\documentclass[aps,prl,twocolumn]{revtex4}

\usepackage{graphicx}

\begin{document}
\title{Comment on ``Dynamical Foundations of Nonextensive Statistical Mechanics"}
\author{A.~M. Crawford}
\author{N. Mordant}
\email{nm88@cornell.edu}
\author{E. Bodenschatz}
\affiliation{LASSP, Cornell University, Ithaca NY 14853, USA}
\author{A. M. Reynolds}
\affiliation{Silsoe Research Institute, Wrest Park, Silsoe, Bedford, MK45 4HS, UK}
\maketitle

In a recent letter~\cite{beck}, Christian Beck described a theoretical link between a family of stochastic differential equations and the probability density functions (PDF) derived from the formalism of nonextensive statistical mechanics. He applied the theory  to explain experimentally measured PDFs from fully developed fluid turbulence, in particular the PDF of one acceleration component $a$ of  fluid particles. He found that the PDF
\begin{equation}
p(a)=C/(1+\frac{1}{2}\beta (q-1) a^2)^{\frac{1}{q-1}}
\end{equation}
with $q=\frac{3}{2}$, $\beta=4$ and $C=2/\pi$ (for normalization to variance 1) captured the experimental data \cite{nature,jfm} reasonably well.

Here we show that equation~(1) is in  disagreement with experimental observations and conclude that Beck's  application of Tsallis statistics to Lagrangian turbulence is not justified.

Figure 1 shows a comparison of the experimental data with the prediction of  Eq. 1 from the letter (dashed line). Clearly, the theoretical  prediction does not correctly capture the functional form of the PDF.  This is most clearly seen when the contributions $a^4P(a)$ to the fourth order moment are considered. More generally, the log-log plot shows that the experimental  data does not support the  power law decay predicted by Beck.

As in the earlier experiment~\cite{nature,jfm} (the data used by Beck in~\cite{beck}), we used  silicon strip detectors to follow the motion of 45 $\mu m$ diameter tracer particles at 70,000 measurements per second at a Taylor based Reynolds number $R_\lambda=690$.
The flow  and the  other experimental conditions were
the same as earlier~\cite{nature,jfm}.
To remove measurement noise, the trajectories were lowpass-filtered with a gaussian kernel of width $0.17\tau_\eta$ \cite{PhysicaD}. The total number of data points gathered was  $1.7\times10^8$.
\begin{figure}[!htb]
\includegraphics[width=7cm]{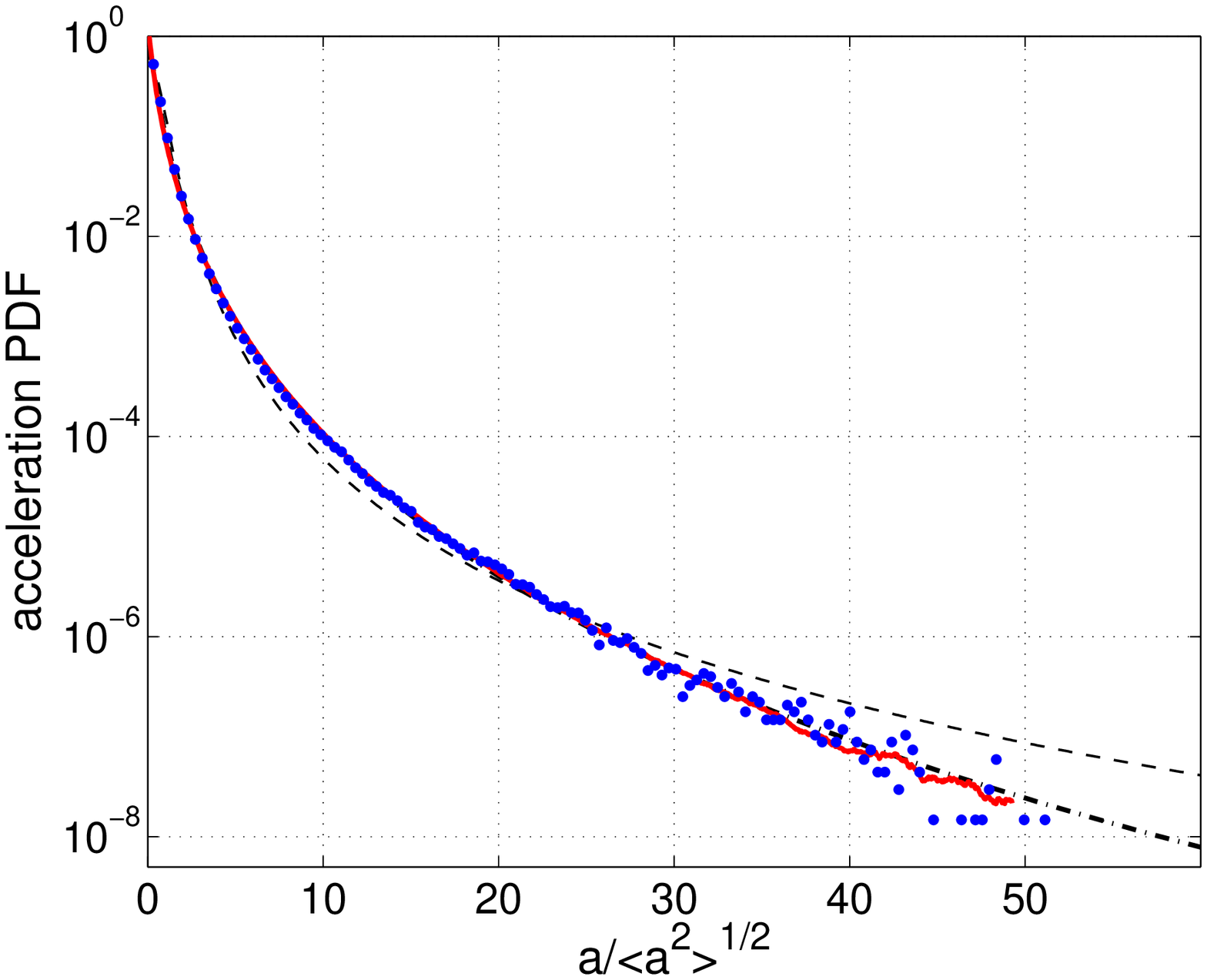}\\
\includegraphics[width=8cm]{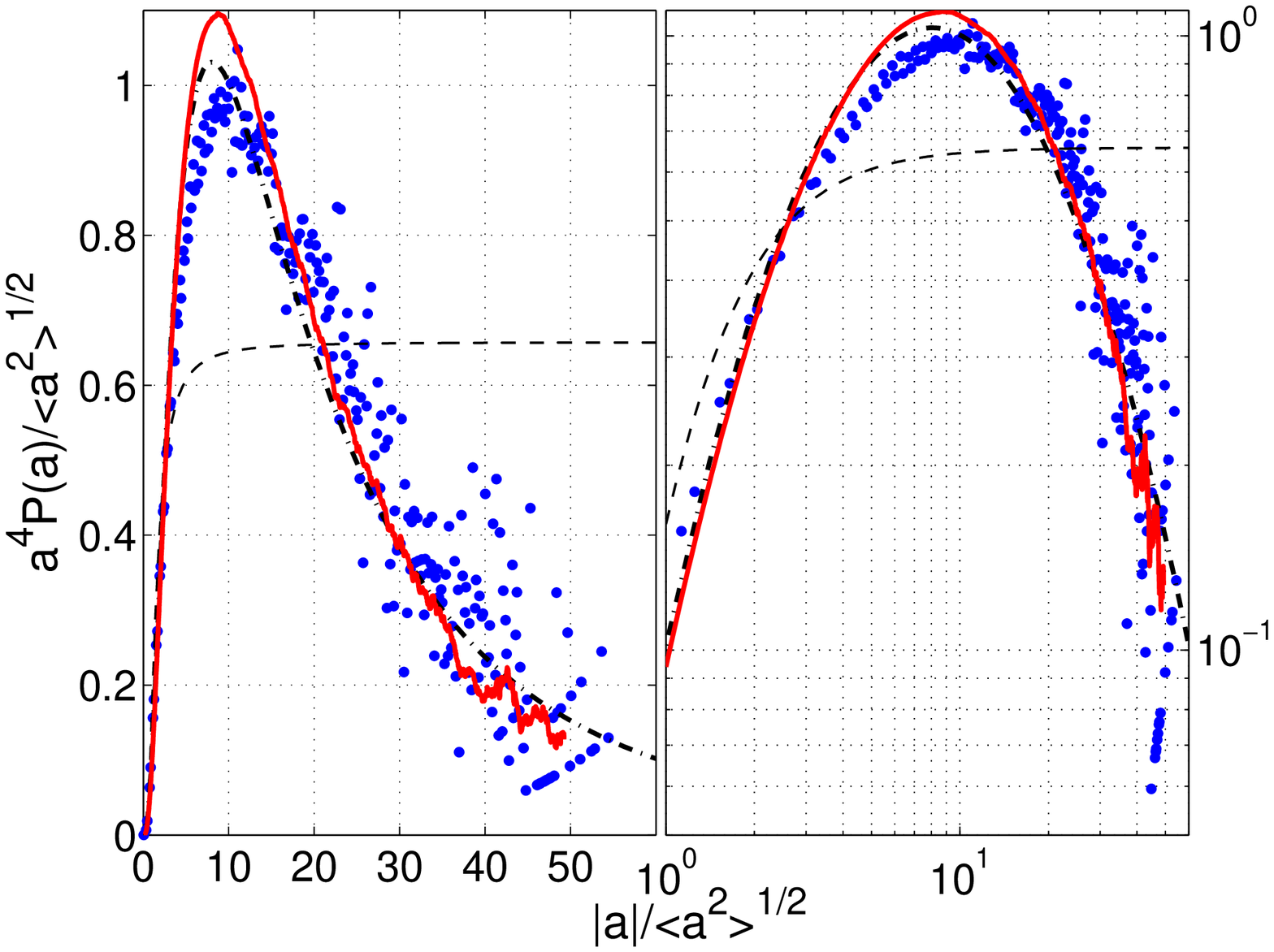}
\caption{Probability density function of fluid particle acceleration $P(a)$. Dots: data points. Dashed line: Beck's model for $q=3/2$. Dot-dashed: Beck's log-normal model for $s^2=3$. Line: Reynolds' model at $R_\lambda=690$ and filtered at the same scale as our data.
Bottom: $a^4P(a)$ vs $|a|$, in linear and loglog scale.}
\end{figure}

Recent theoretical developments by Reynolds~\cite{rey} and Beck \cite{beck1,becklog} also suggest that the Tsallis statistics is not adequate to decribe the Lagrangian acceleration and that generalizations are needed to fit  the data. These new approaches fit our new  data reasonably  well~\cite{PhysicaD}, however, it needs to be stressed that the Beck's lognormal model~\cite{becklog} requires a fit to our data to determine free parameters. The only theory that uses parameters  from independent experimental and numerical results is that by Reynolds~\cite{rey}. His theory uses log-normal statistics and is designed  to describe the full Lagrangian time dynamics and is not restricted  to the dissipation time scale as, for example, in the letter by Beck. Reynolds' model gives excellent agreement with the experimental observations, as shown in Fig.~1 by the solid line.
\acknowledgments{This work is supported by NSF-PHY9988755.}

 \end{document}